\documentclass[journal]{IEEEtran}
\usepackage[utf8]{inputenc}
\usepackage{amsmath,amssymb,graphicx,booktabs,array,multirow,xcolor}
\usepackage[numbers,sort]{natbib}
\setcitestyle{citesep={], [\kern-\fontdimen2\font}}
\usepackage{tikz}\usetikzlibrary{calc,positioning,arrows.meta,fit,shapes.symbols,backgrounds}
\usepackage{hyperref}
\usepackage{pdfpages}
\begin{document}
\title{Directed Brain--Body Graphs Across Timescales: Rate-Matched Granger Causality With Per-Edge Robustness Values}
\author{Lu\'{\i}s~A.~Alexandre and Karolina~Finc%
\thanks{This work has been submitted to the IEEE for possible publication. Copyright may be transferred without notice, after which this version may no longer be accessible. The supplementary material follows the references.}%
\thanks{L.~A.~Alexandre was supported by UID/04516/NOVA Laboratory for Computer Science and Informatics (NOVA LINCS) with the financial support of FCT.IP. K.~Finc was supported by the Excellence Initiative -- Research University programme awarded to the Nicolaus Copernicus University in Toru\'{n}.}%
\thanks{L.~A.~Alexandre is with NOVA LINCS, Universidade da Beira Interior, 6201-001 Covilh\~{a}, Portugal (e-mail: lfbaa@ubi.pt).}%
\thanks{K.~Finc is with the Institute of Advanced Studies, Centre for Modern Interdisciplinary Technologies, Nicolaus Copernicus University in Toru\'{n}, 87-100 Toru\'{n}, Poland (e-mail: finc@umk.pl).}}
\bstctlcite{IEEEexample:BSTcontrol}
\maketitle
\begin{abstract}
\textit{Objective:} Brain and body signals are sampled at different rates, and unmeasured common influences can make one signal
appear to drive another. We aimed to estimate directed brain--body graphs without a common sampling grid and to quantify each edge's
robustness to hidden drivers. \textit{Methods:} Rate-matched Granger causality (RMGC) predicts each signal at a resolution matched to
its sampling rate, from causal windows of the other signals' own samples. Each edge receives a robustness value: the strength of the
weakest simulated common driver, in a fixed class, that reproduces it. We compared RMGC with established methods on a simulated
benchmark (64 replicates) and on sleep (CAP Sleep Database) and awake recordings (BBBD, MultiPhysio-HRC), with breathing observed and
hidden. \textit{Results:} RMGC matched the best established method's recall (0.92) at less than half the false-positive rate of a
common-grid model (0.05 against 0.12), and declared a reversed edge in 4\% of replicates, against 100\% for that model. Robustness
values never overstated the driver behind a spurious edge and separated true from spurious edges better than edge size (area under
the curve 0.80 against 0.40). Of brain--heart edges stable with breathing hidden, 49\% in sleep and 94\% and 79\% awake could be
reproduced by a hidden driver no stronger than breathing. \textit{Conclusion:} RMGC estimates directed graphs from multirate signals
and events with controlled false positives and per-edge robustness values. \textit{Significance:} RMGC lets brain--body studies
report how strong a shared driver such as breathing would need to be to explain each directed edge.
\end{abstract}


\section{Introduction}
\IEEEPARstart{T}{he} brain continuously receives signals from the body while contributing to the regulation of cardiac,
respiratory, and autonomic activity, and these exchanges matter for perception and behavior \citep{azzalini2019}. They also matter
clinically: an imbalanced brain--heart interaction has a negative impact on health \citep{silvani2016}, and in patients with
disorders of consciousness, combining information on cardiac cycle modulations with electroencephalography (EEG) markers improved
the classification of individual patients \citep{raimondo2017}. The quantification of brain--heart interplay has therefore been
proposed as a source of biomarkers for changes in cognitive, emotional, and autonomic state \citep{catrambone2021}. Network
physiology \citep{ivanov2021} studies such coordination as relationships among interacting systems, and sleep studies have linked
changes in brain--heart--respiratory connectivity to transitions between sleep stages \citep{bashan2012,bartsch2015}.

One way to describe these relationships is the direction of prediction: whether the past of one signal improves the prediction of
another beyond its own history and the other observed signals. Granger causality formalizes this question \citep{granger1969},
and related analyses have characterized brain--heart networks during sleep \citep{faes2014}. Predictive direction alone does not
establish a direct physiological pathway \citep{barrett2013} or the effect of an intervention \citep{eichler2010}, and
its estimates depend on two properties of physiological recordings.

The first is the coexistence of very different sampling rates and timescales. EEG is sampled hundreds of times per second, other
sensors much more slowly, and heartbeats form a sequence of events; a fast signal can also carry slow information, such as the
power of a rhythm. Placing all signals on one grid distorts this: coarse bins turn an indirect path into an apparently direct one,
a fine grid needs many past samples for slow dependencies, and filling in a slow signal between its samples either delays it or,
with interpolation, carries information from the future. Measurement noise, filtering, and the relation between sampling intervals
and interaction delays all change which directed dependencies are detected \citep{nalatore2007,barnett2011filter,barnett2017subsample}.

The second is shared influences absent from the model. Respiration affects both cardiac variability and neural activity
\citep{zelano2016}, and omitting it changes the inferred direction of cardiovascular relationships \citep{porta2012respiration}; a
common influence reaching two signals at different delays makes one predict the other without any direct interaction
\citep{granger1969}. Volume conduction similarly creates spurious connectivity between EEG channels \citep{haufe2013}. An observed
edge therefore raises a second question alongside its predictive strength: how strong would an unmeasured common driver have to be
to produce it?

Existing methods address parts of these problems. Mixed-frequency Granger tests use series observed at different rates
\citep{ghysels2016granger}, multiscale formulations quantify interactions at several temporal resolutions \citep{faes2017multiscale},
and causal structures have been learned from subsampled or mixed-rate series \citep{tank2019subsampled,assaad2022entropy,zhu2025mind}.
Neural methods extend the range of nonlinear relationships \citep{tank2022ngc,cheng2023cutsplus} but model the signals
on a single grid, and causal discovery with latent confounders \citep{gerhardus2020lpcmci} reasons about hidden variables without
quantifying their strength per edge. Sensitivity analysis in regression expresses the strength an omitted variable would need, using
measured variables as benchmarks \citep{cinelli2020}. We know of no method that combines prediction at each signal's own rate, edges
across timescales, and a per-edge measure of robustness to unmeasured common drivers.

Here we introduce rate-matched Granger causality (RMGC), which predicts each signal at a resolution matched to its sampling rate
while other signals enter through windows of their own past samples, and EEG sources also through their power. Each edge receives a
robustness value, found by searching for a simulated common driver that reproduces the edge after its fitted contribution has been
removed, and expressed relative to a measured driver such as respiration. A neural extension (RMGC-NN) tests for nonlinear edges.
The paper makes three contributions: a Granger causality test that uses each signal at its own rate, including event series; a
per-edge robustness value expressed in units of a measured driver; and an evaluation against established methods given the same
information, on a benchmark with known ground truth and in overnight sleep recordings and two awake datasets, with placebo and
surrogate controls.

\section{Methods}
\subsection{Signals, Edges, and Tests}\label{sec:setting}
We observe $p$ signals, each as samples at its own rate or as event times (heartbeats). The nodes of the graph are defined before the
analysis: sampled signals (EEG derivations or regional averages, breathing, skin conductance), heart rate (the rate of each completed beat, $60/\mathrm{RR}$, RR: the beat interval in seconds, at the time of the beat), and, in experiments, a stimulus, which can drive other signals but is never
predicted. No node may be a linear combination of others, and deriving a node must not shift its timing (Section~S2).

A directed edge $u\to v$ is declared when the past of $u$ improves the least-squares prediction of $v$ given the past of $v$ and of all
other signals. Each row of the regression matrix $\mathbf X$ contains regressors built only from data before the prediction instant:
an intercept, $v$'s own past, and one block of regressors per source (Section~\ref{sec:rmgc}). With $\mathrm{RSS}$ the residual sum
of squares of the full model and $\mathrm{RSS}_{-u}$ that of the model without $u$'s block, the \emph{size} of the edge is the
partial $R^2$,
\begin{equation}
a = 1-\mathrm{RSS}/\mathrm{RSS}_{-u},
\end{equation}
the share of the variance left unexplained without $u$ that $u$'s block explains. Its significance is judged by the conditional
Granger $F$ test \citep{geweke1982,lutkepohl2005},
\begin{equation}
F = \frac{a/q_u}{(1-a)/(n-k)},
\end{equation}
with $q_u$ columns in $u$'s block, $n$ prediction instants, and $k$ columns in $\mathbf X$. The result is a summary Granger causality
graph \citep{eichler2007}: an edge states that $v$ depends on $u$'s past within the span of its windows (at most 16~s), without
resolving the lag. Every ordered pair is a candidate edge (the stimulus is never a target), and edges are declared from the pairs'
$p$-values by the Benjamini--Hochberg procedure \citep{bh1995} at a false discovery rate of 0.05 over the ordered pairs of one run (one
replicate, segment, or recording). Because tests on the same recording are dependent, this 0.05 is approximate.

\subsection{Rate-Matched Granger Causality}\label{sec:rmgc}
RMGC defines three regular time grids, or \emph{levels}, with steps $\Delta$ of 20~ms, 200~ms, and 2~s (fast, medium, slow). An interval $(t_k,t_k+\Delta]$ is a \emph{bin}, and a signal's value in a bin is the mean of its samples in it. Each signal is predicted
only at its \emph{home level}, the fastest level not faster than its sensor, bin by bin; heart rate is predicted at the medium level,
as the rate of the latest completed beat. The target's own past enters as its previous 15, 20, and 8 bins at the three levels (0.3, 4,
and 16~s), spans fixed a priori to cover each level's dynamics.

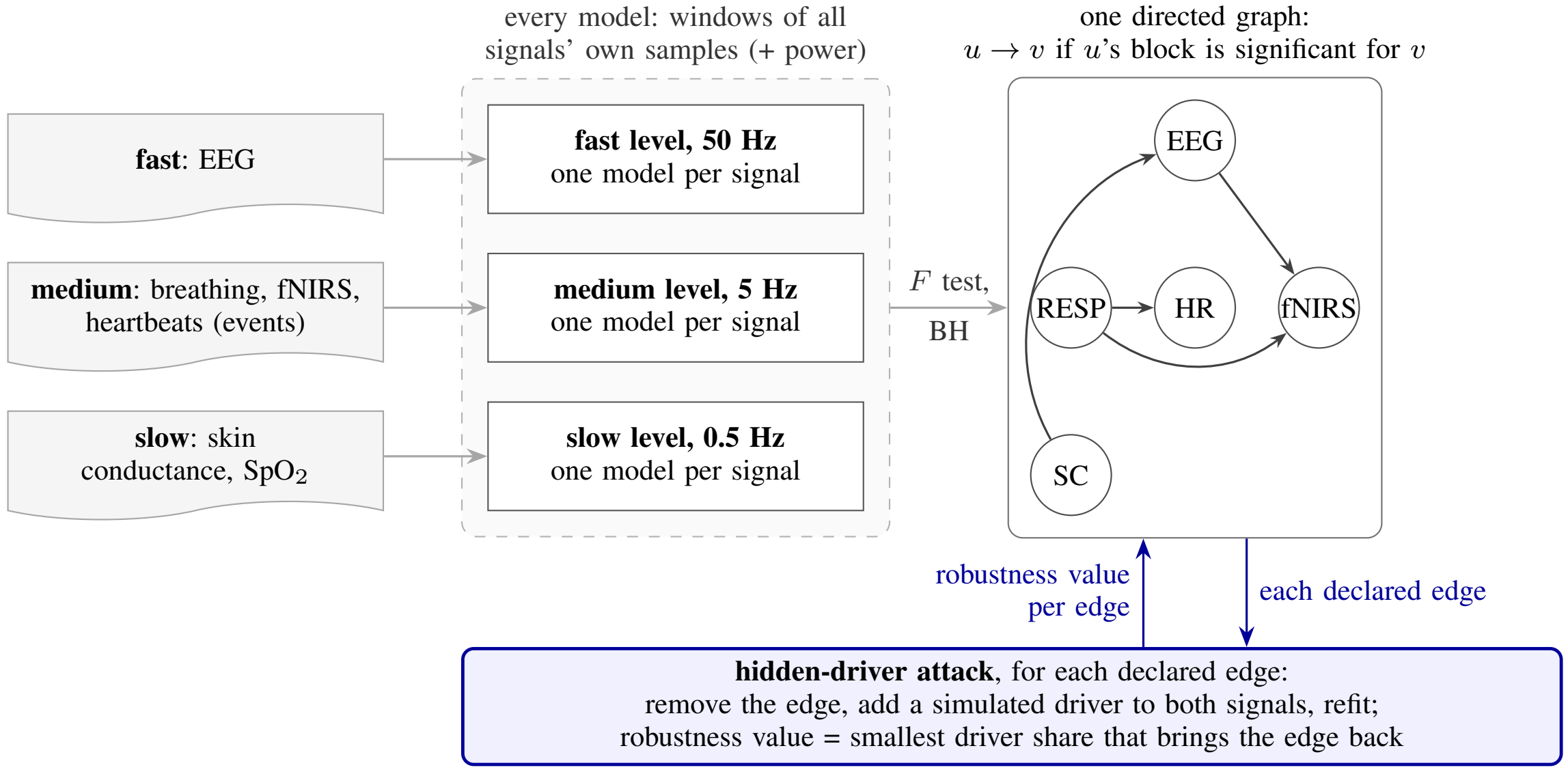
\begin{figure*}[t]\centering
\begin{tikzpicture}[font=\footnotesize,
  data/.style={tape, tape bend top=none, draw=gray!70, fill=gray!8, text width=3.4cm, minimum height=1.05cm, align=center, execute at begin node={\hyphenpenalty=10000}},
  model/.style={rectangle, draw=black!65, fill=white, text width=3.4cm, minimum height=1.05cm, align=center},
  sig/.style={circle, draw=black!70, fill=white, minimum size=0.78cm, inner sep=0pt, font=\footnotesize},
  att/.style={draw=blue!60!black, fill=blue!6, rounded corners=3pt, minimum height=0.95cm, align=center, line width=0.9pt},
  arr/.style={-{Stealth[length=2mm]}, gray!70, line width=0.6pt},
  edg/.style={-{Stealth[length=1.6mm]}, black!75, line width=0.6pt},
  aarr/.style={-{Stealth[length=2mm]}, blue!60!black, line width=0.6pt}]
\node[data] (d0) {\textbf{fast}: EEG};
\node[data, below=0.55cm of d0] (d1) {\textbf{medium}: breathing, fNIRS,\\heartbeats (events)};
\node[data, below=0.55cm of d1] (d2) {\textbf{slow}: skin conductance, SpO$_2$};
\node[model, right=1.0cm of d0] (m0) {\textbf{fast level, 50 Hz}\\one model per signal};
\node[model, right=1.0cm of d1] (m1) {\textbf{medium level, 5 Hz}\\one model per signal};
\node[model, right=1.0cm of d2] (m2) {\textbf{slow level, 0.5 Hz}\\one model per signal};
\foreach \i in {0,1,2} \draw[arr] (d\i) -- (m\i);
\begin{scope}[on background layer]
  \node[draw=gray!60, dashed, rounded corners=4pt, fill=gray!4, inner xsep=7pt, inner ysep=7pt, fit=(m0)(m2),
        label={[align=center, font=\footnotesize, text=gray!40!black]above:every model: windows of all\\signals' own samples (+ power)}] (models) {};
\end{scope}
\coordinate (g) at ($(models.east)+(2.6,0)$);
\node[sig] (EEG)  at ($(g)+(0.35,1.62)$) {EEG};
\node[sig] (RESP) at ($(g)+(-0.85,0)$) {RESP};
\node[sig] (HR)   at ($(g)+(0.35,0)$) {HR};
\node[sig] (FN)   at ($(g)+(1.55,0)$) {fNIRS};
\node[sig] (SC)   at ($(g)+(-0.85,-1.62)$) {SC};
\draw[edg] (RESP) -- (HR);
\draw[edg] (RESP) to[bend right=38] (FN);
\draw[edg] (EEG) -- (FN);
\draw[edg] (SC) to[bend left=50] (EEG);
\begin{scope}[on background layer]
  \node[draw=black!55, rounded corners=4pt, fill=white, inner sep=6pt, fit=(EEG)(RESP)(FN)(SC)(HR),
        label={[align=center, font=\footnotesize]above:one directed graph:\\$u\to v$ if $u$'s block is significant for $v$}] (graph) {};
\end{scope}
\draw[arr] (models.east) -- node[above, font=\footnotesize, text=gray!40!black]{$F$ test,} node[below, font=\footnotesize, text=gray!40!black]{BH} (graph.west);
\node[att, anchor=north west, text width=10.4cm] at ($(models.west |- graph.south)+(0,-1.05)$) (attack)
  {\textbf{hidden-driver attack}, for each declared edge:\\
   remove the edge, add a simulated driver to both signals, refit;\\
   robustness value = smallest driver share that brings the edge back};
\coordinate (gd) at ($(graph.south)+(0.5,0)$);
\coordinate (gu) at ($(graph.south)+(-0.5,0)$);
\draw[aarr] (gd) -- node[right, font=\footnotesize, text=blue!60!black]{each declared edge} (gd |- attack.north);
\draw[aarr] (gu |- attack.north) -- node[left, font=\footnotesize, text=blue!60!black, align=right]{robustness value\\per edge} (gu);
\end{tikzpicture}
\caption{RMGC. \emph{Inputs} (left): the signals at their own rates, grouped by the level that matches their sampling rate.
\emph{Models} (middle): each signal is predicted at its level by its own least-squares model; every model uses causal windows of all
signals' own past samples, and the power of EEG sources. \emph{Output} (right): one directed graph, with an edge $u\to v$ when $u$'s
block of regressors is significant for $v$ ($F$ test per source, Benjamini--Hochberg over the pairs), within or across timescales;
the edges shown are an example. The hidden-driver attack (bottom) takes each declared edge and returns its robustness value. RESP: breathing; HR: heart rate; fNIRS: functional near-infrared spectroscopy; SC: skin conductance; SpO$_2$: oxygen saturation; BH: Benjamini--Hochberg.}\label{fig:arch}
\end{figure*}

Every other signal $u$ enters through a block of regressors computed from its samples as recorded, without interpolation onto the
target's grid. For a target bin starting at $t_k$, window $j$ is $W_j=[t_k-e_{j+1},\,t_k-e_j)$, with edges in multiples of the
target's bin width,
\begin{equation}
e_j = j\Delta\ (j=0,\dots,8),\quad e_{9}=16\Delta,\ e_{10}=32\Delta,\ \dots
\end{equation}
doubling while $e_j\le 16$~s. The first eight windows have the target's bin width; further back the window edges double, so the windows are
fine near the present and coarse further back, and reach 10--16~s with few regressors. Windows narrower than a source's sampling
interval are merged with the next. Window $j$ contributes the mean of $u$'s samples, $f_j$ (heart rate: the mean of the beat rates in it; stimulus: the number of stimulus events; an empty window takes the latest value). An EEG source entering a slower target contributes a second value per window, the logarithm of
the variance of its samples, $g_j$: this \emph{power channel} carries fast-to-slow coupling, where a slow signal follows the strength
of a rhythm rather than its waveform. The values $f_j$ (and $g_j$) of one source form its block, tested with all other blocks and the
target's own past kept. With $v(t_k)$ the target's value in the bin starting at $t_k$, and $f_{u,j}$, $g_{u,j}$ source $u$'s window values for that bin,
the model is
\begin{equation}
v(t_k)=c+\sum_{l=1}^{L}\phi_l\,v(t_{k-l})+\sum_{u\neq v}\sum_{j}\big(\beta_{u,j}f_{u,j}+\gamma_{u,j}g_{u,j}\big)+\varepsilon_k,
\end{equation}
with $L=15$, 20, or 8 at the three levels and $\gamma_{u,j}=0$ for sources without a power channel; testing $u$'s block tests that all its $\beta_{u,j}$ and $\gamma_{u,j}$ are zero. Fig.~\ref{fig:arch} shows the design; Section~S1 discusses aliasing and the alternative
of representing sources at the target's bins (bin pooling).

\subsection{Nonlinear Extensions}\label{sec:nonlinear-main}
The quadratic variant adds the squares of the standardized source summaries to the same $F$ test. RMGC-NN keeps every linear test
and adds an out-of-sample test, by two-fold cross-fitting and a one-sided Diebold--Mariano test, of whether a small network on the
source's summaries improves the prediction of the target's residuals; an edge is declared on linear or nonlinear evidence, with a
Bonferroni factor of two (Section~S6.1). RMGC is the default.

\subsection{Robustness Value}\label{sec:attack}
For a declared edge $u\to v$ of size $a$, the robustness value asks how strong a hidden driver would have to be to produce an edge of
that size if $u$ had no influence on $v$. The strength of a driver $H$ added to a signal $x$ as $bH$ is its \emph{variance share},
\begin{equation}
s=\frac{b^2\,\mathrm{Var}(H)}{b^2\,\mathrm{Var}(H)+\mathrm{Var}(x)},
\end{equation}
measured where each signal enters $v$'s model: on $v$'s bins and on $u$'s window values, taking $u$'s window where the driver is
strongest. The attack 1) removes the edge, subtracting the fitted contribution of $u$'s block from $v$; 2) adds a unit-variance
driver to $u$ with delay 0.25~s and to $v$ with a longer delay, so that $u$ receives it first, with shares $s_u$ and $s_v$; and 3)
refits $v$'s model, searching for the smallest strength at which $u$'s block reaches size $a$ again (a coarse scan followed by seven
bisection steps). The driver class has 20 members: persistent noise (Ornstein--Uhlenbeck, 2 or 10~s) and damped oscillators (0.1 and
0.25~Hz), acting linearly with delay differences of 0.5, 1, and 2~s, or through their square or saturation. The \emph{robustness value}
(RV) is the smallest larger share, $\max(s_u,s_v)$, over the drivers that reproduce the edge; it is 1 if none does with a share of at
most 0.95. A low RV means that a weak driver in the class sufficed; the RV is an upper bound, since a driver outside the class might
need less (Section~S1).

Following \citet{cinelli2020}, the RV is compared with a measured driver, breathing, whose shares $B_u$ and $B_v$ are defined in the
same places: the largest $R^2$ of $u$'s window values on breathing's windows, and the $R^2$ of $v$'s bins on them, in the model with
breathing. The \emph{breathing multiple} $k=\mathrm{RV}/\max(B_u,B_v)$ expresses the RV in units of breathing's strength; $k\le 1$
means that a hidden driver as strong as breathing could produce the edge.

\subsection{Evaluation}\label{sec:benchmark-main}
\subsubsection{Benchmark} A simulator with known ground truth (Section~S3) generates seven signals at the rates of real
sensors: a stimulus (STIM), two EEG-like signals (EEG1, EEG2; 250~Hz), breathing (RESP; 25~Hz), heartbeats giving heart rate (HR), a functional near-infrared spectroscopy-like signal (fNIRS; 8~Hz), and skin conductance (SC; 4~Hz). Most latent nodes follow first-order stochastic differential equations with time constant $\tau$ and Wiener process $W$,
$dx=\tau^{-1}\big(-x+\sum_j c_j h_j(y_j(t-\delta_j))\big)dt+\sqrt{2/\tau}\,dW$, integrated with a 4-ms step, where $y_j$ are the
node's parents, with strengths $c_j$ and delays $\delta_j$, and $h_j$ is the identity, the parent's power envelope, or its square. The six true edges are STIM$\to$EEG1 (50~ms), EEG1$\to$EEG2 (20~ms), SC$\to$EEG1
(0.5~s), EEG2$\to$fNIRS through EEG2's power (2~s), RESP$\to$HR (0.5~s), and RESP$\to$fNIRS (1~s).  Eight scenarios
(clean; hidden arousal and hidden breathing, each mild and strong; a stimulus tempo near heart rate; volume conduction; a squared edge)
have eight 10-min replicates each; all but the clean one include a hidden arousal node that drives HR and SC. They contain traps: edges faked by hidden drivers, the reversed edge HR$\to$RESP and the indirect
edge SC$\to$EEG2 that common grids create, and mixing between EEG sensors. Five success criteria were set for the benchmark: 1) stimulus, fast, and
medium recall at least that of vector autoregressions on common grids; 2) cross-scale recall above 0.5; 3) a false-positive rate
of at most 0.05 in the clean scenario; 4) robustness values that rank faked edges below true ones (AUC of at least 0.8); and 5) a
split-half Jaccard index of at least 0.75 between the edges declared on the two halves of a recording.

\subsubsection{Comparisons} The central comparison gives established methods the same information as RMGC: a vector autoregression on
common grids at 50 and 2~Hz (grid VAR), multiscale Granger causality \citep{faes2017multiscale}, and PCMCI (PC with momentary conditional independence tests) \citep{runge2019pcmci}, each
with the EEG power columns and one test per ordered pair. The methods as published, including neural Granger causality and LPCMCI (latent PCMCI),
are compared in Section~S4. Mixed-data sampling regressions and mixed-frequency Granger tests \citep{ghysels2016granger,gotz2016mfvar} were not compared directly (Section~S1). Three ablations remove
the power channel, rate matching (every signal interpolated to 50~Hz), and the windows (bin pooling). Calibration was checked in 64
replicates of a null scenario without any edge; the share of replicates with a false edge is given with a 95\% Wilson interval and
compared with 5\% by a one-sided binomial test.

\subsubsection{Recordings} We used three datasets (Section~S2): the six healthy-subject recordings with breathing of the Cyclic Alternating Pattern (CAP) Sleep Database \citep{terzano2001,goldberger2000}, cut into 10-min segments by sleep stage (two EEG derivations and breathing, about 128 and 25~Hz after averaging, heart rate, and oxygen saturation in one night); experiments 4 and 5 of the Brain, Body, and Behavior Dataset (BBBD) \citep{madsen2026bbbd},
43 and 48 participants watching short educational videos, with six recordings of 2.4--6.5~min each, released at 128~Hz (four regional EEG averages, breathing, heart rate, pupil size, and electrooculography); and MultiPhysio-HRC \citep{bussolan2025}, recorded during human--robot collaboration (HRC), our stress test: 52 participants, recorded at 256~Hz with a 12-electrode dry headset during rest, guided breathing, cognitive,
virtual-reality, and manual and robot-assisted tasks (four regional EEG averages, breathing, heart rate, skin conductance, and trapezius electromyography; segments of at least 2~min, median 4.1~min).
Heartbeats were the hand-corrected R-peaks released with BBBD and, in the other two datasets, R-peaks detected on the ECG (Section~S2.4); segments and recordings that failed quality criteria were skipped.
All three are publicly available and no new data were collected. The BBBD experiments were approved by the Institutional Review
Boards of the City University of New York \citep{madsen2025bbbdpreprint}, and MultiPhysio-HRC was collected in accordance with the
Declaration of Helsinki with the approval of the Ethics Committee of the University of Applied Sciences and Arts of Southern
Switzerland \citep{bussolan2025}; in both, informed consent was obtained from all participants. The CAP Sleep Database is distributed
on PhysioNet under the Open Data Commons Attribution License. Secondary analysis of these public, de-identified data required no further ethical review.

In every segment RMGC was run with breathing and with breathing hidden (the \emph{hide-breathing test}). Edges present only without
breathing are \emph{breathing-sensitive}, which suggests that breathing contributes to them; edges present in both runs are
\emph{stable}. An edge counted as breathing-sensitive only if its $p$-value in the run without breathing passed the Benjamini--Hochberg threshold of the run with breathing, so that it could not appear merely because the threshold moved. As controls, a
\emph{placebo}, a copy of breathing shifted by half the recording, was added to the model and hidden in the same way; breathing$\to$heart rate served as a
positive control; and surrogates without coupling (another recording's EEG; in the awake data also the body signals shifted circularly
against the EEG) gave background rates of brain--body edges. Other intervals are 95\% percentile bootstrap intervals
\citep{efron1993} (2000 resamples of replicates within scenarios, or of recordings or participants).

\section{Results}
\subsection{Detection and Calibration on the Benchmark}\label{sec:res-detect}
\begin{figure}[t]\centering\includegraphics[width=\columnwidth]{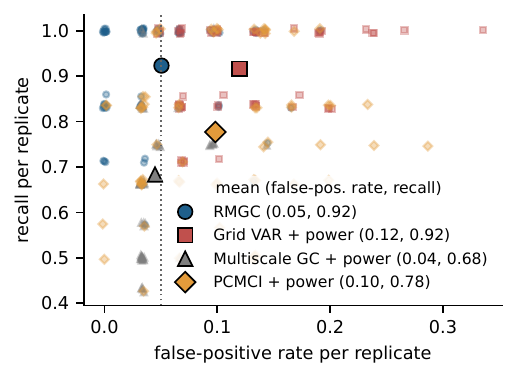}
\caption{Recall against false-positive rate in each of the 64 benchmark replicates, for RMGC and the established methods given the
same EEG power features and one test per pair. Large markers: means. Dotted line: a false-positive rate of 0.05.}\label{fig:benchmark}\end{figure}

\begin{figure*}[t]\centering\includegraphics[width=\textwidth]{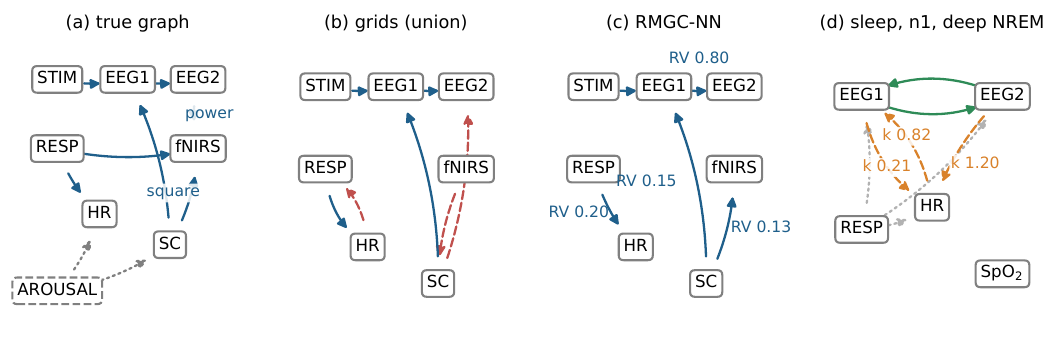}
\caption{Example graphs. Solid blue, true edges, and dashed red, false edges (b, c); solid green, stable edges, and dashed orange, breathing-sensitive edges (d); gray dotted, edges from breathing (d) and from the hidden arousal (a). (a) True graph of the
nonlinear-edge scenario. (b) Edges declared by the union of common grids in one replicate, and (c) by RMGC-NN in the
same replicate, with robustness values. (d) One segment of deep non-rapid-eye-movement (NREM) sleep of recording n1: edges from breathing in the run with breathing,
and the edges of the hidden-breathing run with their breathing multiples $k$.}\label{fig:graphs}\end{figure*}

Given the same power features and one test per pair (Table~\ref{tab:onetest}, Fig.~\ref{fig:benchmark}), RMGC reached a recall of
0.92 at a false-positive rate of 0.05. The grid VAR matched its recall at 2.4 times its false-positive rate (0.12) and declared the
reversed edge HR$\to$RESP in every replicate and the indirect edge SC$\to$EEG2 in every replicate outside volume conduction. Multiscale
Granger causality had a similar false-positive rate (0.04 against 0.05) but a much lower recall (0.68); PCMCI fell between (0.78 and 0.10). RMGC declared the reversed edge in 4\% and the indirect edge in 2\% of the replicates. With its history limited to the grid VAR's 4~s, RMGC kept these advantages (recall 0.90, false-positive rate 0.057, the two edges in 2\% and 0\%; Section~S4.6). As published, without power features, the
established methods reached recalls of 0.43--0.77 (Section~S4). The nonlinear variants matched RMGC's recall and
false-positive rate and found the squared edge (RMGC-NN in 8 of 8 replicates of the nonlinear scenario, the quadratic variant in 6 of 8; Fig.~\ref{fig:graphs} shows example graphs). In paired tests, each difference stated here had a Holm-adjusted $p<0.001$ (Section~S5.4). RMGC met success criteria 1--3 and 5 (split-half Jaccard index 0.80 on average, at least 0.75 in six of the eight scenarios; Section~S4); criterion 4 is reported below.

\begin{table}[t]\centering\caption{Matched comparison on the benchmark (64 replicates): recall, false-positive rate (FPR), detection AUC (area under the receiver operating characteristic curve), and shares of replicates declaring the reversed edge HR$\to$RESP and the indirect edge SC$\to$EEG2. All methods use the EEG power columns and one test per ordered pair; GC: Granger causality.}\label{tab:onetest}\footnotesize\setlength{\tabcolsep}{3.5pt}
\begin{tabular}{@{}lccccc@{}}\toprule
 & Recall & FPR & AUC & HR$\to$RESP & SC$\to$EEG2$^*$\\\midrule
Grid VAR + power & 0.92 & 0.12 & 0.96 & 1.00 & 1.00\\
Multiscale GC + power & 0.68 & \textbf{0.04} & 0.92 & \textbf{0.00} & 1.00\\
PCMCI + power & 0.78 & 0.10 & 0.93 & 0.10 & 0.77\\\midrule
RMGC & 0.92 & 0.05 & \textbf{0.98} & 0.04 & 0.02\\
RMGC, quadratic & \textbf{0.93} & 0.05 & \textbf{0.98} & 0.04 & \textbf{0.00}\\
RMGC-NN & \textbf{0.93} & 0.05 & \textbf{0.98} & 0.04 & 0.02\\\bottomrule
\end{tabular}
\\[3pt]\parbox{\columnwidth}{\footnotesize $^*$Outside the volume-conduction scenario.}
\end{table}

Removing any one component lowered at least one measure (Table~\ref{tab:ablation}; $p\le 0.008$ for each change below, Table~S19). Without the power channel, cross-scale recall fell from 0.93 to 0.49 and the
fast-to-slow edge EEG2$\to$fNIRS was found in 2\% of the replicates instead of 94\%. Interpolating every signal to 50~Hz brought the
reversed edge back in every replicate and raised the false-positive rate to 0.12. Bin pooling lowered the overall recall slightly (0.90) and medium-scale recall from 0.80 to 0.68. In 64 null replicates, RMGC declared a false edge in one (1.6\%; 0.3--8.3\%), consistent with
5\%, as were its variants (6 and 5) and multiscale Granger causality (6); the grid VAR (8; 6.5--22.8\%) and PCMCI (23; 25.3--48.2\%)
exceeded it (Section~S5).

\begin{table}[t]\centering\caption{Ablations of RMGC (64 replicates): recall, FPR, medium-scale and cross-scale recall, and share of replicates declaring the reversed edge HR$\to$RESP}\label{tab:ablation}\footnotesize\setlength{\tabcolsep}{3.5pt}
\begin{tabular}{@{}lccccc@{}}\toprule
 & Recall & FPR & Medium & Cross-scale & HR$\to$RESP\\\midrule
Full RMGC & \textbf{0.92} & \textbf{0.05} & \textbf{0.80} & 0.93 & 0.04\\
Without power & 0.75 & \textbf{0.05} & 0.77 & 0.49 & 0.02\\
Interpolated, 50~Hz & 0.71 & 0.12 & 0.60 & 0.50 & 1.00\\
Bin pooling$^*$ & 0.90 & \textbf{0.05} & 0.68 & \textbf{0.94} & \textbf{0.00}\\\bottomrule
\end{tabular}
\\[3pt]\parbox{\columnwidth}{\footnotesize $^*$Seven scenarios (not run on the nonlinear one).}
\end{table}

\subsection{Validity and Discrimination of the Robustness Value}\label{sec:res-rv}
\begin{figure*}[t]\centering\includegraphics[width=\textwidth]{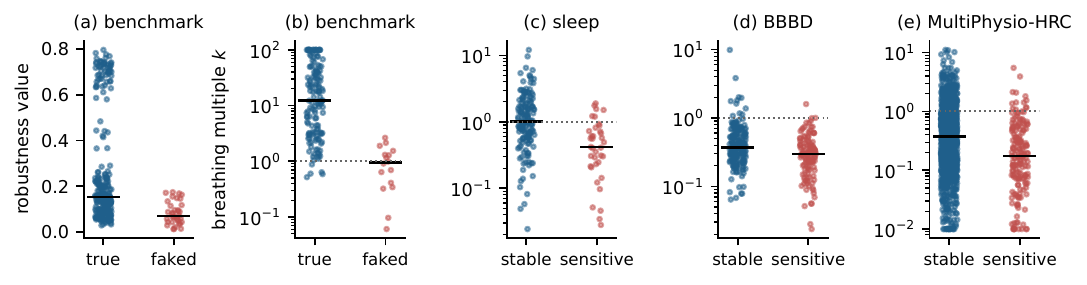}
\caption{(a) Robustness values of true and faked edges in the benchmark (RMGC). (b) Breathing multiples of the same edges where
breathing was observed; dotted line: as strong as breathing. (c) Sleep, (d) BBBD and (e) MultiPhysio-HRC: breathing multiples of the
brain--heart edges of the runs without breathing, stable and breathing-sensitive. Black bars: medians.}\label{fig:robust}\end{figure*}

Across the 64 replicates, RMGC declared 265 true edges and 36 edges faked by a hidden driver. Every faked edge had an RV at most the
actual share of the driver that produced it (36 of 36; 28 of 28 for RMGC-NN). The RV separated true from faked edges with an area under the receiver operating characteristic curve (AUC) of 0.80
[0.74, 0.85], against 0.40 for edge size (Fig.~\ref{fig:robust}(a)). Criterion 4 (AUC of at least 0.8) was narrowly missed by RMGC, whose unrounded AUC was 0.7996, and met by RMGC-NN (0.805), so RMGC-NN met all five criteria and RMGC four (Table~S20). The AUC difference between the RV and edge size was 0.40 [0.31, 0.49] ($p<0.001$). Under strong hidden arousal, every faked edge was larger than every true edge.
Relative to breathing (Fig.~\ref{fig:robust}(b)), on the 141 true and 16 faked edges where breathing was observed, the breathing multiple
separated them with AUC 0.94 [0.92, 0.96]; $k\le 1$ for 56\% of the faked edges and 4\% of the true ones. For two drivers outside the
searched class, the RV stayed below the actual share; a slow drift's faked edges were still separated from true ones (AUC 0.90), but an
amplitude-modulating oscillation's faked edge looked as robust as true edges (AUC 0.36).

\subsection{Real Recordings}\label{sec:res-real}
Table~\ref{tab:empirical} summarizes the controls. The positive control was declared in 93\% of the sleep segments, 99\% of the BBBD
recordings, and 59\% of the MultiPhysio-HRC segments (82\% during guided breathing). Hiding breathing made new edges appear far more
often than hiding the placebo in sleep (38 against 5) and BBBD (369 against 148), and 670 against 424 in MultiPhysio-HRC.
Surrogates gave background rates of brain--heart edges of 4\% in sleep and 2\% in BBBD, against 34\% and 7\% in the real recordings;
in MultiPhysio-HRC the background was 12\%, against 28\%.

\begin{table}[t]\centering\caption{Empirical controls: positive control (breathing$\to$heart rate declared; analyzed segments or recordings in parentheses), new edges when breathing or the placebo is omitted, and brain--heart edges in surrogates without coupling and in the real recordings, as shares of the possible ones}\label{tab:empirical}\footnotesize\setlength{\tabcolsep}{3.5pt}
\begin{tabular}{@{}lcccc@{}}\toprule
 & Positive & Omitted: & \multicolumn{2}{c}{Brain--heart}\\\cmidrule(l){4-5}
 & control & resp./placebo & surrogates & real\\\midrule
Sleep (115) & 93\% & 38 / 5 & 4\% & 34\%\\
BBBD (518) & 99\% & 369 / 148 & 2\% & 7\%\\
MultiPhysio-HRC (557) & 59\% & 670 / 424 & 12\% & 28\%\\\bottomrule
\end{tabular}

\end{table}

Table~\ref{tab:brain_heart} compares brain--heart edges (EEG$\leftrightarrow$heart rate) with one definition. In sleep,
breathing-sensitive edges needed a median of 0.42 times breathing's share and stable ones 1.01; about half of the stable edges (49\%)
could be produced by a hidden driver as strong as breathing, and the breathing multiple separated the two groups with AUC 0.75, about as
well as edge size (0.70). In both awake datasets, stable brain--heart edges needed a median of 0.37 (BBBD) and 0.38
(MultiPhysio-HRC) times breathing's share, and 94\% and 79\% of them had $k\le 1$. Of the edges RMGC-NN added, 3 of 6 in sleep,
64\% in BBBD, and 55\% in MultiPhysio-HRC were breathing-sensitive, and most were fragile (Section~S6).

\begin{table}[t]\centering\caption{Brain--heart edges (EEG$\leftrightarrow$heart rate) of the runs without breathing: stable and breathing-sensitive edges, their median breathing multiple $k$, the shares with $k\le 1$, and how well $k$ and edge size separate the two groups (AUC)}\label{tab:brain_heart}\footnotesize\setlength{\tabcolsep}{3.5pt}
\begin{tabular}{@{}lcccc@{}}\toprule
 & Edges & Median $k$ & $k\le 1$ & AUC\\
 & stab./sens. & stab./sens. & stab./sens. & $k$ / size\\\midrule
Sleep & 152 / 37 & 1.01 / 0.42 & 49\% / 84\% & 0.75 / 0.70\\
BBBD & 179 / 123 & 0.37 / 0.30 & 94\% / 98\% & 0.63 / 0.68\\
MultiPhysio-HRC & 1038 / 171 & 0.38 / 0.17 & 79\% / 92\% & 0.65 / 0.54\\\bottomrule
\end{tabular}

\end{table}

\section{Discussion}
\subsection{Rate Matching}
The common-grid failures in the benchmark follow from placing signals with different rates and timescales on one grid. A coarse grid
turns an indirect path into a direct edge, a fine grid needs too many lags for slow relations, and held or interpolated samples
reverse edges. This agrees with the known dependence of Granger causality on filtering and on the relation between sampling interval
and interaction delay \citep{barnett2011filter,barnett2017subsample}. Predicting each signal at its own level, with sources entering
through windows of their own samples, removes these failures with ordinary least squares and $F$ tests. Mixed-frequency Granger
tests need regular sampling in a fixed ratio of frequencies \citep{ghysels2016granger}, and multiscale Granger causality rescales series that share one grid \citep{faes2017multiscale}. State-space Granger causality \citep{barnett2015ss} also needs one common grid, and for Gaussian processes transfer entropy equals Granger causality \citep{barnett2009te}, so neither removes the need for rate matching. RMGC adds signals sampled at any rate or recorded as events, a graph over
many signals with multiple-testing control, and explicit edges across timescales.

The ablations separate the contributions: interpolation onto one grid brought the reversed edge back into RMGC itself, the power
channel carried the fast-to-slow edge, and the windows kept medium-scale recall. The ablations do not isolate the component that avoids the indirect edge (Section~S4.5). The power channel matches physiology: in neurovascular and brain--body coupling, slow signals follow the power of a neural rhythm
rather than its voltage. With the same power features and the same 4-s history, RMGC kept its lower false-positive rate and rarely declared the reversed and indirect edges, so its advantage comes from not resampling rather than from its longer history.

\subsection{Robustness Values}
No method avoided the edges produced by strong hidden drivers, and edge size was a misleading guide, because slow, persistent
drivers fake large edges with a small variance share. The RV addresses this as sensitivity analysis does in regression
\citep{cinelli2020}: the useful question is not whether an edge could be confounded, but whether a confounder as strong as a known
one could explain it. Causal discovery with latent confounders reasons about hidden variables
\citep{gerhardus2020lpcmci}; the RV states the strength they would need for each edge, and the breathing multiple expresses it in
physiological units.

 Its AUC was 0.80 in the benchmark; that of the breathing multiple was 0.75 in sleep and 0.63--0.65 awake, below edge size in BBBD
(Table~\ref{tab:brain_heart}). The RV is an upper bound on the weakest driver. A low RV shows that a weak driver suffices; a high RV shows only that no driver in the
class did with less, and a driver that acts in ways the class cannot mimic, such as amplitude modulation, can make a faked edge look robust.

\subsection{Breathing in Brain--Heart Networks}
Our results bear on whether directed edges in the brain--heart networks described during sleep
\citep{bashan2012,bartsch2015,faes2014} reflect neural control. In sleep, about half of the stable brain--heart edges (49\%) could be
produced by a hidden driver of breathing's strength, so they should not be read as direct neural control of the heart without
further evidence. 

Awake, brain--heart edges were more fragile: stable ones needed only about 0.4 times breathing's share in both datasets
(Table~\ref{tab:brain_heart}). The two datasets differ in task, movement, and background rate and gave nearly the same median, which suggests that
the fragility does not come from recording quality alone.  Segment length did not set the background rate (BBBD's equally short recordings had a low one), so circular-shift surrogates are
an inexpensive check for any new dataset.

On the same MultiPhysio-HRC segments RMGC found the known breathing$\to$heart-rate edge more often than the established methods, especially
during guided breathing (82\% against 32--68\%; Section~S7). It also declared more breathing-sensitive edges: the
sensitivity that finds slow couplings also picks up slow common drivers, which is why per-edge robustness values are needed. 

\subsection{Limitations and Future Work} Edges are predictive, not interventional. Subtracting the heartbeat-locked ECG template, the cardiac field artifact, from the sleep EEG kept 94\% of the brain--heart edges and Table~\ref{tab:brain_heart}'s sleep row nearly unchanged (Section~S7.4); the awake datasets were not checked. The benchmark is linear apart from the power
and squared edges, has two hidden drivers and was calibrated by hand, and its strong scenarios were set to trap the established methods, and it lacks the artifacts and nonstationarity of real recordings. The $F$ tests assume independent residuals of constant variance, which heart rate held between beats and autocorrelated physiological residuals violate; error control held in null simulations containing both, but in real data surrogate background rates are an empirical diagnostic, not the false-discovery rate of the real graphs,
and they exceeded the nominal level in MultiPhysio-HRC, where short nonstationary segments violate the tests' assumptions.

RVs are point estimates without intervals, and upper bounds over a class of 20 drivers. The sleep analysis rests on six recordings with two EEG derivations each.
RMGC treats each recording separately, and group-level statements rest on prevalence across participants (Section~S7). Stable edges are not ground
truth. The established methods ran on real data only on subsets fixed in advance (Section~S7).

Future work includes thresholds calibrated on surrogates, bootstrap intervals for RVs, datasets with controlled stimuli, and learned spatial filters as EEG nodes to reduce volume conduction.

\section{Conclusion}
Rate-matched Granger causality estimates directed graphs from signals recorded at different rates or as events, without a common
sampling grid. On a benchmark with known ground truth, given the same information as established methods, it matched the highest
recall with less than half the false-positive rate of a common-grid model, and it rarely declared the reversed and indirect edges
that resampling creates. Its error control held in null simulations. Its robustness value, the strength a hidden common driver would need to produce an edge, never overstated the driver behind a spurious edge and separated true from spurious edges where edge size failed. Relative to measured breathing, it showed that about half of the stable brain--heart edges in
sleep, and most of them in two awake datasets, could be produced by a hidden driver no stronger than breathing. Directed brain--body
relationships should therefore be reported together with their robustness to shared drivers.

\section*{Code and Data Availability}
The code will be made publicly available upon acceptance and is available to reviewers on request. The CAP Sleep Database is
available on PhysioNet \citep{goldberger2000}, BBBD \citep{madsen2026bbbd} and MultiPhysio-HRC \citep{bussolan2025data} on Zenodo.

\section*{Acknowledgment}
Claude (Anthropic) wrote most of the analysis and figure code under the authors' direction, drafted text, and edited the language. 
The authors designed the study, wrote the text, reviewed and tested all code (which generates
every reported number, table, and figure from the stored results), checked every statement and reference against its source, and
take full responsibility for the content.

\bibliographystyle{IEEEtranN}
\bibliography{references}
\onecolumn
\includepdf[pages=-]{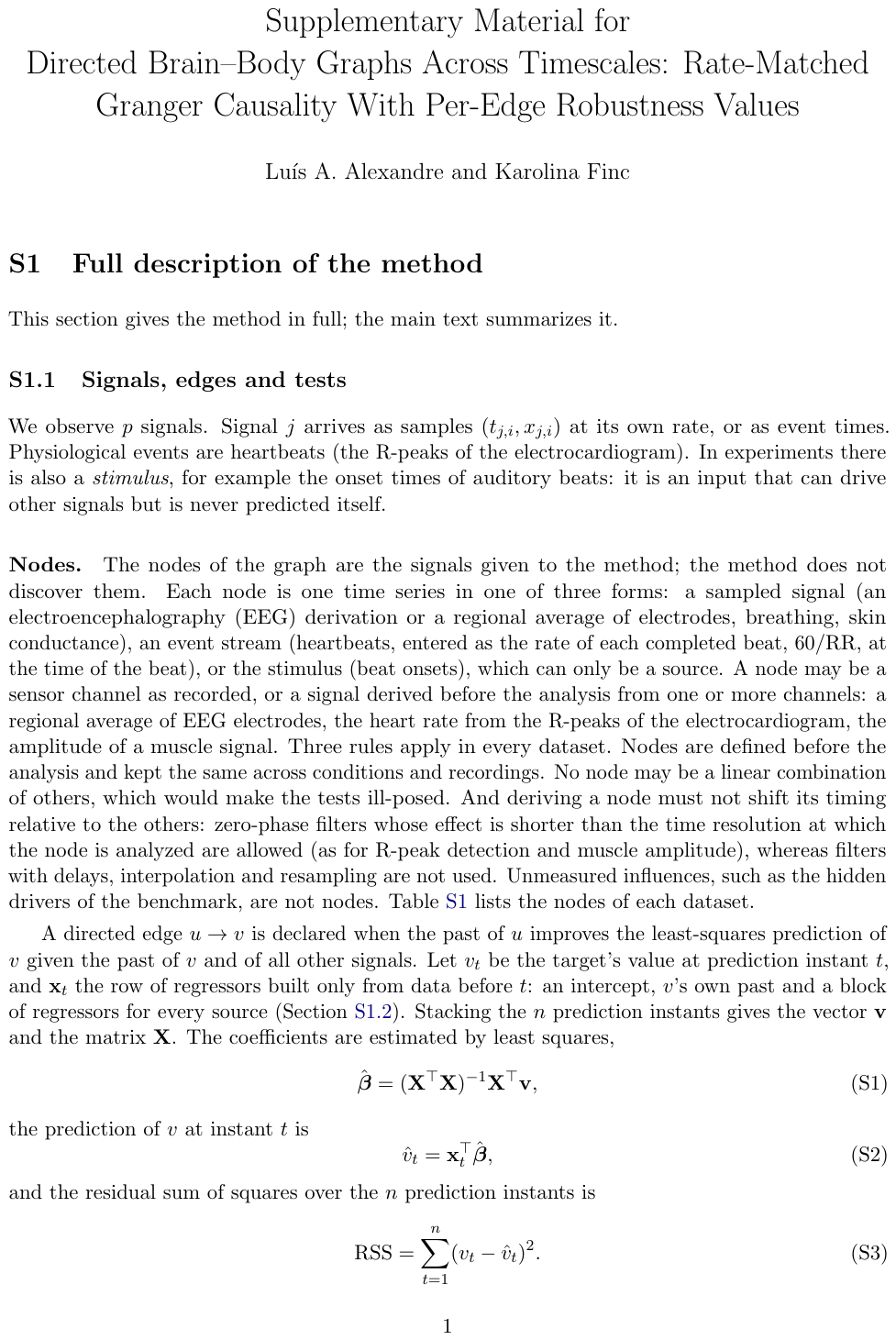}
\end{document}